\documentstyle[12pt]{article}

\newlength{\dinwidth}
\newlength{\dinmargin}
\def\lapproxeq{\lower .7ex\hbox{$\;\stackrel{\textstyle
<}{\sim}\;$}}
\def\gapproxeq{\lower .7ex\hbox{$\;\stackrel{\textstyle
>}{\sim}\;$}}

\begin{document}
\titlepage
\begin{flushright}
DTP/95/34  \\
April 1995 \\
\end{flushright}

\begin{center}
\vspace*{2cm}
{\Large \bf BFKL predictions at small $x$ from $k_T$ \\

\bigskip
and collinear factorization viewpoints}

\vspace*{1cm}
J.\ Kwieci\'{n}ski\footnote{On leave from the Henryk
Niewodnicza\'{n}ski Institute of Nuclear Physics, 31-342
Krak\'{o}w, Poland.} and A.\ D.\ Martin,

Department of Physics, University of Durham, Durham, DH1 3LE,
England \\

\end{center}

\vspace*{3cm}
\begin{abstract}
Hard scattering processes involving hadrons at small $x$ are
described by a $k_T$-factor- \linebreak ization formula driven by
a BFKL gluon.  We explore the equivalence of this description to
a
collinear-factorization approach in which the anomalous
dimensions $\gamma_{gg}$ and $\gamma_{qg}/\alpha_S$ are expressed
as power series in $\alpha_S \log (1/x)$, or to be precise
$\alpha_S/\omega$ where $\omega$ is the moment index.  In
particular we confront the collinear-factorization expansion with
that extracted from the BFKL approach with running coupling
included.
\end{abstract}

\newpage

Recently there have been several studies [1-5] of the validity
and possible modification of the conventional
Altarelli-Parisi (or GLAP) description of deep inelastic
scattering in the small $x$ region that has become accessible at
HERA, $x \sim 10^{-4}$.  The relevant modifications are the
inclusion of contributions which are enhanced by powers of $\log
(1/x)$, but which lie outside the leading (and next-to-leading)
Altarelli-Parisi perturbative expansion.  Formally they
correspond to the expansion of the anomalous dimensions
$\gamma_{gg}$ and $\gamma_{qg}/\alpha_S$ as
power series in $\alpha_S/\omega$ where $\omega$ is the moment
index.  An alternative approach which automatically resums {\it
all} these leading $\log (1/x)$ contributions to $\gamma_{gg}$
and $\gamma_{qg}/\alpha_S$ is provided by the BFKL equation
coupled with the $k_T$-factorization formula for calculating
observable quantities \cite{KT,AKMS2}.  The main aim of this
paper is to explore the connection between these two approaches.
To be specific we study the relation between the
collinear-factorization formula with $\log (1/x)$ terms included
and the $k_T$-factorization formula based on the solution of the
BFKL equation \cite{BFKL} with running coupling $\alpha_S$.  We
show that both approaches generate the same first few terms in
the perturbative expansion of $\gamma_{gg}$ and, more important,
of $\gamma_{qg}$, which are presumably the most relevant
contributions for the description of deep inelastic scattering in
the HERA range.  They differ substantially, however, in the
asymptotically small $x$ regime.

Deep inelastic unpolarised electron-proton scattering may be
described in terms of two structure functions, $F_2 (x,Q^2)$ and
$F_L (x, Q^2)$.  As usual, the kinematic variables are defined to
be $Q^2 = -q^2$ and $x = Q^2/2p.q$, where $p$ and $q$ are the
four-momenta of the incoming proton and virtual photon probe
respectively.  At small values of $x, x \lapproxeq 10^{-3}$,
these observables reflect the distribution of gluons in the
proton, which are by far the dominant partons in this kinematic
region.  The precise connection between the small $x$ structure
functions and the gluon distribution is given by the
$k_T$-factorization formula \cite{KT,AKMS2},
\begin{equation}
F_i (x, Q^2) = \int \frac{dk_T^2}{k_T^2} \int_x^1
\frac{dx^\prime}{x^\prime} F_i^{\gamma g} \biggl
(\frac{x}{x^\prime}, k_T^2, Q^2 \biggr ) f(x^\prime, k_T^2)
\label{a1}
\end{equation}
\noindent with $i = 2, L$, which is displayed pictorially in
Fig.\ 1.  The gluon distribution $f(x, k_T^2)$, unintegrated over
$k_T^2$, is a solution of the BFKL equation, while $F_i^{\gamma
g}$ are the off-shell gluon structure functions which at
lowest-order are determined by the quark box (and crossed-box)
contributions to photon-gluon fusion, see Fig.\ 1.

For sufficiently large values of $Q^2$ the leading-twist
contribution is dominant, and it is most transparent to discuss
the $Q^2$ evolution of $F_i (x, Q^2)$ in terms of moments.  Then
the $x^\prime$ convolution of (\ref{a1}) factorizes to give
\begin{equation}
\overline{F}_i (\omega, Q^2) = \int \frac{dk_T^2}{k_T^2}
\overline{F}_i^{\gamma g} (\omega, k_T^2, Q^2) \overline{f}
(\omega, k_T^2)
\label{a2}
\end{equation}
\noindent where the moment function
\begin{equation}
\overline{f} (\omega, k_T^2) \equiv \int_0^1 \frac{dx}{x} \:
x^\omega f(x, k_T^2),
\label{a3}
\end{equation}
\noindent with similar relations for $\overline{F}_i$ and
$\overline{F}_i^{\gamma g}$.

\bigskip
\noindent {\large \bf Fixed $\alpha_S$ :  $k_T$-factorization to
collinear-factorization}

It is illuminating to first consider the case of fixed coupling
$\alpha_S$ \cite{CH}.  Then the photon-gluon moments,
$\overline{F}_i^{\gamma g}$ are simply functions of $\tau \equiv
Q^2/k_T^2$ (and $\omega$), for massless quarks.  Hence (\ref{a2})
becomes a convolution in $k_T^2$ which, in analogy with the
$x^\prime$ convolution, may be factorized by taking moments a
second time.  In this way we obtain representations for the
$\overline{F}_i$ with factorizable integrands
\begin{equation}
\overline{F}_i (\omega, Q^2) = \frac{1}{2 \pi i} \int_{c - i
\infty}^{c + i \infty} d\gamma \: \tilde{F}_i^{\gamma g} (\omega,
\gamma) \: \tilde{f} (\omega, \gamma)(Q^2)^\gamma
\label{a4}
\end{equation}
\noindent with $c = {\textstyle \frac{1}{2}}$.  The (double)
moments $\tilde{F}_i^{\gamma g}$ and $\tilde{f}$ of the gluon
structure functions and the gluon distribution are respectively
defined by
\begin{eqnarray}
\tilde{F}_i^{\gamma g} (\omega, \gamma) & = & \int d \tau \:
\tau^{-\gamma-1} \: \overline{F}_i^{\gamma g} (\omega, \tau)
\label{a5}
\\
\tilde{f}(\omega, \gamma) & = & \int dk_T^2 \:
(k_T^2)^{-\gamma-1} \:
\overline{f} (\omega, k_T^2)
\label{a6}
\end{eqnarray}
\noindent where the $\tilde{F}_i^{\gamma g}$ are dimensionless,
but $\tilde{f}$ carries the dimension $(k_0^2)^{-\gamma-1}$.
Representation (\ref{a4}) enables the leading-twist contribution
to be identified from a knowledge of the analytic properties of
$\tilde{f}$ and $\tilde{F}_i^{\gamma g}$ in the complex $\gamma$
plane.

The gluon distribution $f (x, k_T^2)$ satisfies the BFKL
equation, which in moment space has the form
\begin{equation}
\overline{f} (\omega, k_T^2) = f^0 (\omega, k_T^2) +
\frac{\overline{\alpha}_S}{\omega} \int \frac{dk_T^{\prime
2}}{k_T^{\prime 2}} K(k_T^2, k_T^{\prime 2}) \overline{f}
(\omega, k_T^{\prime 2})
\label{a7}
\end{equation}
\noindent where $\overline{\alpha}_S \equiv 3\alpha_S/\pi$ and
$K$ is the usual BFKL kernel.  The double-moment function
$\tilde{f}$ is therefore given by
\begin{equation}
\tilde{f} (\omega, \gamma) = \frac{\tilde{f}^0 (\omega,
\gamma)}{1 - (\overline{\alpha}_S/\omega) \tilde{K}(\gamma)}
\label{a8}
\end{equation}
\noindent where $\tilde{K} (\gamma)$ is the eigenvalue of the
BFKL kernel corresponding to the eigenfunction proportional to
$(k_T^2)^\gamma$.  It can be shown that
\begin{eqnarray}
\tilde{K}(\gamma) & = & 2 \Psi (1) - \Psi (1 - \gamma) - \Psi
(\gamma) \nonumber \\
& = & \frac{1}{\gamma} \biggl [ 1 + \sum_{n = 1}^\infty 2 \zeta
(2n + 1) \gamma^{2n + 1} \biggr ]
\label{a9}
\end{eqnarray}
\noindent where $\Psi$ is the logarithmic derivative of the Euler
gamma function, $\Psi (z) \equiv \Gamma^\prime (z)/\Gamma (z)$,
and $\zeta(n)$ is the Riemann zeta function.

We see from (\ref{a4}) that the large $Q^2$ behaviour of
$\overline{F}_i (\omega, Q^2)$ is controlled by the pole at
$\gamma = \overline{\gamma}$ of $\tilde{f} (\omega, \gamma)$ of
(\ref{a8}) which lies to the left of, and nearest to, the contour
of integration in the $\gamma$-plane.  For a physically
reasonable choice of input $\tilde{f}^0$, this pole arises from
the zero of the denominator of (\ref{a8}).  That is
\begin{equation}
\frac{1}{1 - (\overline{\alpha}_S/\omega) \: \tilde{K}(\gamma)}
\: = \: \frac{\gamma R}{\gamma - \overline{\gamma}},
\label{a10}
\end{equation}
\noindent where from (\ref{a9}) we have
\begin{eqnarray}
\overline{\gamma} & = & \frac{\overline{\alpha}_S}{\omega} + 2
\zeta(3) \left ( \frac{\overline{\alpha}_S}{\omega} \right )^4 +
2 \zeta (5) \left ( \frac{\overline{\alpha}_S}{\omega} \right )^6
+ {\cal O} \left (\frac{\overline{\alpha}_S}{\omega} \right )^7,
\label{a11}
\\
R & = & \left ( 1 - \frac{\overline{\alpha}_S}{\omega} \: \left .
\frac{d (\gamma \tilde{K})}{d \gamma} \right
|_{\overline{\gamma}} \right )^{- 1} \nonumber \\
& = & 1 + 6 \zeta(3) \left ( \frac{\overline{\alpha}_S}{\omega}
\right )^3 + {\cal O} \left (\frac{\overline{\alpha}_S}{\omega}
\right )^5.
\label{a12}
\end{eqnarray}
\noindent $\overline{\gamma}$ is the leading-twist anomalous
dimension \cite{JAR}.  If we insert the pole of (\ref{a10}) and
(\ref{a8})
into (\ref{a4}), and we close the contour of integration in the
left-half plane, then we obtain the high $Q^2$ behaviour
\begin{equation}
\overline{F}_i (\omega, Q^2) = \tilde{F}_i^{\gamma g} (\omega,
\overline{\gamma}) \:\: \overline{\gamma} R \left (
\frac{\overline{\alpha}_S}{\omega} \right ) \:\: \tilde{f}^0
(\omega, \overline{\gamma})(Q^2)^{\overline{\gamma}}.
\label{a13}
\end{equation}
\noindent Eq.\ (\ref{a13}) is the usual formula for the
factorization of collinear (or mass) singularities written in
moment space.  This becomes more apparent if we express
(\ref{a13}) in the form
\begin{equation}
\overline{F}_i (\omega, Q^2) = C_i^{\gamma g} (\omega,
\overline{\gamma}) \: g (\omega, Q^2)
\label{a14}
\end{equation}
\noindent where
\begin{equation}
C_i^{\gamma g} (\omega, \overline{\gamma}) = \overline{\gamma}
\tilde{F}_i^{\gamma g} (\omega, \overline{\gamma}) \: R \left (
\frac{\overline{\alpha}_S}{\omega} \right )
\label{a15}
\end{equation}
\noindent is the moment of the (process dependent) coefficient
function and
\begin{equation}
g (\omega, Q^2) = (Q_0^2)^{\overline{\gamma}} \: \tilde{f}^0
(\omega, \overline{\gamma}) \left ( \frac{Q^2}{Q_0^2} \right
)^{\overline{\gamma}}
\label{a16}
\end{equation}
\noindent is the moment function of the (integrated) gluon
density.  Thus we can identify $(Q_0^2)^{\overline{\gamma}} \:
\tilde{f}^0 (\omega, \overline{\gamma})$ with the moment of the
gluon distribution at the \lq\lq starting" scale $Q_0^2$ of
the evolution in $Q^2$.

The quantity $R$, the residue in (\ref{a10}), is renormalisation
scheme dependent \cite{CH}.  For studies of the BFKL equation,
(\ref{a7}), it is appropriate to regularise $R$ by choosing an
inhomogeneous term of the form
\begin{equation}
f^0 (\omega, k_T^2) = G^0 (\omega) \delta (k_T^2 - \mu^2).
\label{b16}
\end{equation}
On the other hand $\overline{\gamma}$, and $\tilde{F}_i^{\gamma
g}$, are scheme independent (at least in the so-called regular
schemes).  The scheme dependence of $R$ is compensated by
subleading contributions of \linebreak ${\cal O}
(\alpha_S(\alpha_S/\omega)^n)$ in the anomalous dimension
$\gamma_{gg}$, which are still at present unknown.  This
cancellation takes place when we allow the coupling to run.

The above collinear factorization formula (\ref{a13}) is true as
it stands for $F_L$, but some care is needed for $F_2$.  First we
check its validity for $F_L$.  Since
$\overline{F}_L^{\overline{\gamma} g} \rightarrow$ constant for
large $\tau$, we see from (\ref{a5}) that $\tilde{F}_L^{\gamma g}
(\omega, \gamma) \sim 1/\gamma$.  However, this potential
singularity at $\gamma = 0$ is cancelled by the $\gamma$ factor
in the numerator of (\ref{a10}).  On the other hand
$\overline{F}_2^{\gamma g} \rightarrow \log (\tau)$ and hence
$\tilde{F}_2^{\gamma g} (\omega, \gamma) \sim
1/\gamma^2$, where the double pole reflects the collinear
singularity associated with the $g \rightarrow q\overline{q}$
transition.  The integration contour in (\ref{a4}) therefore also
encloses the pole
\begin{equation}
\tilde{F}_2^{\gamma g} (\omega, \gamma) \: \tilde{f} (\omega,
\gamma) \sim \frac{1}{\gamma^2} \gamma \sim \frac{1}{\gamma},
\label{c16}
\end{equation}
which gives rise to a \lq\lq scaling sea" contribution to $F_2$
which is independent of $Q^2$.  To remove this contribution and
to focus attention on the effects of the perturbative pole at
$\gamma = \overline{\gamma}$ we consider the observable $\partial
F_2/\partial \log Q^2$, rather than $F_2$ itself.  In this case
the collinear factorization formula is of the form
\begin{equation}
\frac{\partial \overline{F}_2 (\omega, Q^2)}{\partial \log Q^2} =
\sum_q \: 2 e_q^2 \: P_{q g} (\omega, \overline{\gamma}) \: g
(\omega, Q^2)
\label{a17}
\end{equation}
\noindent where the coefficient, or gluon-quark splitting,
function is given by
\begin{equation}
P_{q g} (\omega, \overline{\gamma}) = \overline{\alpha}_S
\Phi (\omega, \overline{\gamma}) \: R \left
(\frac{\overline{\alpha}_S}{\omega} \right ),
\label{a18}
\end{equation}
\noindent with
\begin{equation}
\overline{\alpha}_S \Phi (\omega, \gamma) \equiv \gamma^2
\tilde{F}_2^{\gamma g} (\omega, \gamma)
\label{b18}
\end{equation}
\noindent defined to be a regular function at $\gamma = 0$.
Since $\Phi$ is known \cite{CH} in terms of the quark box (and
crossed
box), we can determine the perturbative expansion of $P_{qg}$ by
calculating
\begin{equation}
P_{qg} (\omega, \overline{\gamma}) = \overline{\alpha}_S \left [
\Phi (\omega, 0) + \overline{\gamma} \left . \frac{\partial
\Phi}{\partial \gamma} \right |_{\gamma = 0} + \ldots \right ] R
\label {c18}
\end{equation}
\noindent with $\overline{\gamma}$ and $R$ given by the
perturbative expansions of (\ref{a11}) and (\ref{a12})
respectively.  To be precise we substitute for
$\overline{\gamma}$ in (\ref{c18}) and obtain the perturbative
expansion of $P_{qg}/\alpha_S$ as a power series in
$\overline{\alpha}_S/\omega$.  This provides the recipe to
compute the leading $\log (1/x)$ contribution to $P_{qg}$.

Although we have expanded the observables in a perturbative
series in $\overline{\alpha}_S/\omega$, we should recall that the
small $x$ behaviour of $F_L$ and $\partial F_2/\partial \log Q^2$
is controlled by the singularities of $\overline{F}_i (\omega,
Q^2)$ in the $\omega$ complex plane.  The singularities arise
from $\overline{\gamma}$.  The leading singularity of
$\overline{\gamma}$ is the BFKL branch point at $\omega =
\omega_L = (4 \log 2) \overline{\alpha}_S$.  To see this we note
that the position of the singularity is controlled by the value
of $\tilde{K}(\gamma)$ at its symmetry point, $\gamma =
{\textstyle \frac{1}{2}}$.  We expand $\tilde{K}$ about this
point
\begin{equation}
\tilde{K} (\gamma) = 4 \log 2 + 14 \zeta (3) (\gamma -
{\textstyle \frac{1}{2}})^2 + \ldots,
\label{a19}
\end{equation}
\noindent and determine the leading singularity as the implicit
solution of
\begin{equation}
1 - \frac{\overline{\alpha}_S}{\omega} \tilde{K}
(\overline{\gamma}) = 0,
\label{a20}
\end{equation}
\noindent see (\ref{a8}).  The leading pole of $\tilde{f}
(\omega, \gamma)$, which lies inside the contour of integration
of (\ref{a4}), is at
\begin{equation}
\overline{\gamma} = \frac{1}{2} - \sqrt{\frac{\omega -
\omega_L}{14 \overline{\alpha}_S \zeta (3)}}
\label{a21}
\end{equation}
\noindent where $\omega_L = (4 \log 2) \overline{\alpha}_S$.  In
$x$ space the leading singularity of the anomalous dimension
gives an $x^{-\omega_L}$ behaviour of the gluon distribution at
asymptotically small values of $x$.  On the other hand the
perturbation series in $\overline{\alpha}_S/\omega$, as in
(\ref{a11}), enables the collinear factorization formula to be
used to investigate the approach to the BFKL $x^{- \omega_L}$
form, as $x$ decreases since
\begin{equation}
\sum_{n = 1} c_n \left (\frac{\overline{\alpha}_S}{\omega} \right
)^n \: \rightarrow \: \sum_{n = 1} c_n \overline{\alpha}_S
\frac{(\overline{\alpha}_S \log 1/x)^{n - 1}}{(n - 1) !}.
\label{a22}
\end{equation}
\noindent The collinear factorization formulae give well-defined
perturbative expansions for $F_L$ and \linebreak $\partial
F_2/\partial \log Q^2$ which allow the leading
$\overline{\alpha}_S \log (1/x)$ contributions to be resummed.
We will discuss the implications of the reduction of
$k_T$-factorization to collinear form after we have implemented
the running of $\alpha_S$.

\bigskip
\noindent {\large \bf Running $\alpha_S$ :
collinear-factorization to $k_T$-factorization}

To see the effect of the running of $\overline{\alpha}_S$ we
simply replace
\begin{equation}
\overline{\gamma} (\overline{\alpha}_S, \omega) \: \rightarrow \:
\overline{\gamma} (\overline{\alpha}_S (Q^2), \omega),
\label{a23}
\end{equation}
\noindent and similarly for $R(\overline{\alpha}_S/\omega)$, in
(\ref{a14}) and (\ref{a17}).  The crucial change is in the $Q^2$
evolution factor of $g (\omega, Q^2)$, which becomes
\begin{equation}
\left (\frac{Q^2}{Q_0^2} \right )^{\overline{\gamma}} \:
\rightarrow \: \exp \left ( \int_{Q_0^2}^{Q^2} \frac{dq^2}{q^2}
\overline{\gamma} (\overline{\alpha}_S (q^2), \omega) \right ).
\label{a24}
\end{equation}
\noindent In the small $x$ BFKL limit $\overline{\gamma}$ is
simply a function of the ratio $\overline{\alpha}_S
(q^2)/\omega$, as in the fixed coupling case.  We see immediately
the important role played by the
non-perturbative region.  To illustrate the effect, it is
sufficient to take
\begin{equation}
\overline{\alpha}_S (q^2) = b/\log (q^2/\Lambda^2),
\label{a25}
\end{equation}
\noindent and to write (\ref{a11}) in the form
\begin{equation}
\overline{\gamma} = \sum_{n = 1}^\infty A_n \left
(\frac{\overline{\alpha}_S (q^2)}{\omega} \right )^n
\label{a26}
\end{equation}
\noindent where the coefficients are known (and in particular
$A_1 = 1$ and $A_2 = A_3 = A_5 = 0$).  Then the exponent in
(\ref{a24}) is given by
\begin{equation}
\int_{Q_0^2}^{Q^2} \frac{dq^2}{q^2} \overline{\gamma} =
\frac{b}{\omega} \log \left ( \frac{\overline{\alpha}_S
(Q_0^2)}{\overline{\alpha}_S (Q^2)} \right ) + \frac{b}{\omega}
\sum_{n = 4}^\infty \frac{A_n}{n - 1} \left \{ \left (
\frac{\overline{\alpha}_S (Q_0^2)}{\omega} \right )^{n - 1} \: -
\: \left ( \frac{\overline{\alpha}_S (Q^2)}{\omega} \right )^{n -
1} \right \}.
\label{a27}
\end{equation}
\noindent The first term on the right-hand-side leads to the
usual double-leading-logarithmic (DLL) behaviour of the gluon
distribution $g (\omega, Q^2)$.  The sum in the second term
builds up the BFKL behaviour, and here we see the dominance of
the $\overline{\alpha}_S (Q_0^2)/\omega$ power series evaluated
at the starting scale $Q_0^2$ as compared to the truly
perturbative power series in $\overline{\alpha}_S (Q^2)/\omega$.
In other words the leading singularity in (\ref{a27}) is the BFKL
branch point at $\omega = \overline{\alpha}_S (Q_0^2) 4 \log 2$.
In principle it should be reabsorbed in the starting distribution
$g (\omega, Q_0^2)$, leaving the perturbative contribution which
is controlled by $\overline{\alpha}_S (Q^2)$; yet in practice it
is the full formula (\ref{a27}) which is used
\cite{EKL,EHW,BF,FRT}.

The above representation contains therefore an equivalent
infrared sensitivity to that contained in the direct BFKL
predictions \cite{AKMS2}, but we see that it has been
explicitly isolated in a factorizable form.  By infrared
sensitivity we mean that the leading singularity in the $\omega$
plane is controlled by $\alpha_S (Q_0^2)$ and not by $\alpha_S
(Q^2)$.  The distinction between $Q^2$ and $Q_0^2$ is, of course,
immaterial in the region $(Q^2 \gapproxeq Q_0^2)$ of
applicability of the genuine leading $\log (1/x)$ approximation,
that is $\alpha_S (Q_0^2) \log (Q^2/Q_0^2) \ll 1$, but $\alpha_S
(Q_0^2) \log (1/x) \sim {\cal O}(1)$.

The BFKL corrections to the DLL contribution only enter the
expansion for the anomalous dimension (\ref{a27}) at order
$(\alpha_S/\omega)^4$ and above, whereas for $P_{q g}$ of
(\ref{a19}) it can be shown that all terms $(n = 0, 1, \ldots)$
are present in the expansion $\sum B_n (\alpha_S/\omega)^n$.  For
this reason we expect that the small $x$ behaviour of $\partial
F_2/\partial \log Q^2$ in the HERA regime will be controlled more
by the perturbative expansion of $P_{q g}$ than of
$\overline{\gamma}$.  However, as $x$ decreases the expansion of
$\overline{\gamma}$ will begin to play a dominant role.

The BFKL equation was originally derived for fixed $\alpha_S$.
The correct way to include the running of $\alpha_S$ is not
firmly established.  The procedure usually adopted is to take
$\alpha_S (k_T^2)$ in (\ref{a7}) so that the DLL limit of GLAP
evolution is obtained.  Here we find the perturbative expansion
obtained from this prescription.  We are therefore able to check
the validity of the procedure by comparing with the expansion
obtained from the renormalization group (or collinear
factorization) approach, that is (\ref{a23})-(\ref{a27}).

If we replace the fixed $\alpha_S$ of (\ref{a7}) by $\alpha_S
(k_T^2)$ the BFKL equation becomes
\begin{equation}
\log \left (\frac{k_T^2}{\Lambda^2} \right ) \overline{f}
(\omega, k_T^2) = \log \left ( \frac{k_T^2}{\Lambda^2} \right )
f^0 (\omega, k_T^2) + \frac{b}{\omega} \int \frac{d k_T^{\prime
2}}{k_T^{\prime 2}} K (k_T^2, k_T^{\prime 2}) \overline{f}
(\omega, k_T^2),
\label{a30}
\end{equation}
\noindent which, in terms of the moment variable $\gamma$
conjugate\footnote{For running $\alpha_S$ we choose to
inter-relate dimensionless quantities $\overline{f}
\leftrightarrow \tilde{f}$, whereas for fixed $\alpha_S$ it was
convenient to allow $\tilde{f}$ to carry dimensions, see
(\ref{a6}).} to $k_T^2/\Lambda^2$, reduces to the differential
equation [10-13]
\begin{equation}
- \frac{\partial \tilde{f} (\omega, \gamma)}{\partial \gamma} =
- \: \frac{\partial \tilde{f}^0 (\omega, \gamma)}{\partial
\gamma} + \frac{b}{\omega} \tilde{K} (\gamma) \tilde{f} (\omega,
\gamma).
\label{a31}
\end{equation}
\noindent From the extension of (\ref{a4}) and (\ref{b18}) to
running $\alpha_S$ we see that the $k_T$-factorization gives
\begin{equation}
\frac{\partial F_2 (\omega, Q^2)}{\partial \log Q^2} = \frac{1}{2
\pi i} \int_{\frac{1}{2} - i \infty}^{\frac{1}{2} + i \infty} d
\gamma \: \overline{\alpha}_S (Q^2) \Phi (\omega, \gamma)
\frac{1}{\gamma} \tilde{f} (\omega, \gamma) \left (
\frac{Q^2}{\Lambda^2} \right )^\gamma,
\label{a32}
\end{equation}
\noindent and similarly for $F_L (\omega, Q^2)$, where the
double-moment of the gluon $\tilde{f} (\omega, \gamma)$ is the
solution of (\ref{a31}).  The leading-twist contribution is
controlled by the solution of the homogeneous form of (\ref{a31})
\cite{JK,JCJK}
\begin{equation}
\tilde{f} (\omega, \gamma) = H^0 (\omega) \exp \left [
\frac{b}{\omega} \int_\gamma d \gamma^\prime \tilde{K}
(\gamma^\prime) \right ]
\label{a33}
\end{equation}
\noindent where $H^0 (\omega)$ will eventually have to be fixed
by the starting gluon distribution, $\tilde{f}^0$.

We now inspect the perturbative expansion of (\ref{a32}) and find
that the first few terms are identical to those in the
renormalization group expansion.  To make this identification we
concentrate on the expansion in terms of powers of
$\overline{\alpha}_S (Q^2)/\omega$ which contain the hard scale
$Q^2$.  The non-perturbative contributions can always be absorbed
into a redefinition of $H^0 (\omega)$.

To begin we note that the leading twist contribution is
controlled by the strip $-1 < \gamma < 0$ of the branch cut in
the $\gamma$ plane, where the branch point at $\gamma = 0$ is
generated entirely by the $1/\gamma$ term in $\tilde{K}
(\gamma)$.  We isolate this singularity by introducing the
function
\begin{equation}
\overline{K} (\gamma) \equiv \tilde{K} (\gamma) -
\frac{1}{\gamma}
\label{a34}
\end{equation}
\noindent such that $\overline{K}$ is regular at $\gamma = 0$.
We then insert the result
\begin{equation}
\frac{b}{\omega} \int_\gamma d \gamma^\prime \: \tilde{K}
(\gamma^\prime) = - \frac{b}{\omega} \log \gamma +
\frac{b}{\omega} \int_\gamma d \gamma^\prime \: \overline{K}
(\gamma^\prime)
\label{a35}
\end{equation}
\noindent into (\ref{a33}) and (\ref{a32}), fold the contour
around the cut and evaluate the discontinuity to obtain
\begin{equation}
\frac{\partial F_2 (\omega, Q^2)}{\partial \log Q^2} = - \sin
\left (\frac{\pi b}{\omega} \right ) H^0 (\omega)
\overline{\alpha}_S (Q^2) I + {\rm higher \: twist}
\label{a36}
\end{equation}
\noindent where
\begin{equation}
I \equiv \int_{-1}^0 d \gamma \: \Phi (\omega, \gamma)(-
\gamma)^{-
\frac{b}{\omega} - 1} \: \exp \left (\frac{b}{\omega} \int_\gamma
d \gamma^\prime \: \overline{K} (\gamma^\prime) \right ) \left
(\frac{Q^2}{\Lambda^2} \right )^\gamma.
\label{a37}
\end{equation}
\noindent The integral $I$ has, of course, to be understood in
the sense of an analytic continuation since it diverges at
$\gamma = 0$.  To expand $I$ in a perturbation series we first
change the variable of integration
\begin{equation}
\gamma \: \rightarrow \: -\rho/t \quad \hbox{where} \quad t
\equiv \log (Q^2/\Lambda^2),
\label{a38}
\end{equation}
\noindent then the integral takes the form
\begin{equation}
I = t^{b/\omega} \int_0^t d \rho \: \Phi (\omega, - \rho/t)
\rho^{- \frac{b}{\omega} - 1} \: \exp \left ( \frac{b}{\omega}
\int_{- \rho/t} d \gamma^\prime \: \overline{K} (\gamma^\prime)
\right ) e^{- \rho}.
\label{a39}
\end{equation}
\noindent We use (\ref{a9}) to expand the first exponential
factor in
(\ref{a39})
\begin{equation}
\exp \left ( \frac{b}{\omega} \: \int_{- \rho/t} \: d
\gamma^\prime \: \overline{K} (\gamma^\prime) \right ) = \exp
\left ( \frac{b}{\omega} \: \sum_{n = 1} \: \frac{2 \zeta (2n +
1)}{2n + 1} \left (\frac{\rho}{t} \right )^{2n + 1} \right ),
\label{a40}
\end{equation}
\noindent where we have omitted a factorizable non-perturbative
contribution coming from the upper limit.  When we expand the
exponential in (\ref{a40}) and insert the series into (\ref{a39})
we encounter
integrals of the form
\begin{eqnarray}
\int_0^\infty \: d \rho \: \rho^{- \frac{b}{\omega} + 2n} \: e^{-
\rho} & = & \Gamma \left (- \frac{b}{\omega} + 2n + 1 \right )
\nonumber \\
& = & \Gamma (- b/\omega) \: (- b/\omega)^{2n + 1} (1 + {\cal O}
(\omega)).
\label{a41}
\end{eqnarray}
\noindent Here the contribution from $t < \rho < \infty$ gives
higher-twist terms which vanish as $1/Q^2$, modulo logarithmic
corrections.  The term $\Gamma (- b/\omega)$ can be reabsorbed
into the starting distribution, where it belongs, and we find the
perturbative expansion of $I$ is of the form
\begin{equation}
I \sim t^{b/\omega} \: \Gamma (- b/\omega) \: \Phi (\omega, 0) \:
\left [1 - \frac{b}{\omega} \: \sum_{n = 1} \: \frac{2 \zeta (2n
+ 1)}{2n + 1} \: \left (\frac{\overline{\alpha}_S (Q^2)}{\omega}
\right )^{2n + 1} \right ] + {\rm higher \: order \: terms}.
\label{a42}
\end{equation}
\noindent We see that the first two terms $(n = 1,2)$ are
identical to the first two terms $(n = 4,6)$ in the perturbative
expansion of (\ref{a27}), which are proportional to
$\overline{\alpha}_S (Q^2)^3$ and $\overline{\alpha}_S (Q^2)^5$
respectively.  The DLL contribution in (\ref{a27}) corresponds to
the $t^{b/\omega}$ factor in (\ref{a42}).

To generate the expansion of $P_{qg}/\alpha_S$ as a power series
in $\overline{\alpha}_S/\omega$ we expand the function $\Phi
(\omega, \gamma)$ of (\ref{a39}) around $\gamma = 0$.  This
procedure generates the same first three terms as those in the
expansion shown in (\ref{c18}).  At higher order,
$(\overline{\alpha}_S/\omega)^3$ and above, we see that the terms
of ${\cal O} (\omega)$ in (\ref{a41}), as well as various other
contributions, will also contribute to the expansion of $P_{qg}$.
However, it is the first few terms that are important for the
onset of the BFKL behaviour in the HERA small $x$ regime
\cite{EKL,EHW}.  Note that the perturbative expansion in
(\ref{a42}) contains an additional factor of $b/\omega$ which
enables this series to be separated from the perturbative
expansion of $\Phi (\omega, \gamma)$.

Before we conclude, we can gain further insight into the relation
between the BFKL equation and collinear factorization in the
case of running $\alpha_S$ if we estimate the integral
(\ref{a32}) using the saddle-point method.  Inserting (\ref{a33})
we see that the position of the saddle-point,
$\overline{\gamma}$, is given by the implicit equation
$$
- \frac{b}{\omega} \: \tilde{K} (\overline{\gamma}) \: + \: \log
\left ( \frac{Q^2}{\Lambda^2} \right ) = 0,
$$
\noindent that is by
\begin{equation}
\frac{\overline{\alpha}_S (Q^2)}{\omega} \: \tilde{K}
(\overline{\gamma}) = 1,
\label{a45}
\end{equation}
\noindent which is the same as (\ref{a20}) for fixed $\alpha_S$.
We evaluate the integrand of (\ref{a32}) at $\gamma =
\overline{\gamma}$ and use (\ref{a45}) to rearrange the product
$\tilde{f} (\omega, \overline{\gamma}) \:
(Q^2/\Lambda^2)^{\overline{\gamma}}$ in the form
\begin{equation}
H^0 (\omega) \: \exp \left \{ \left ( \frac{b}{\omega}
\int_{\overline{\gamma}} \: d \gamma^\prime \: \tilde{K}
(\gamma^\prime) \right ) + \overline{\gamma} \: \log \left (
\frac{Q^2}{\Lambda^2} \right ) \right \} = \hat{H}^0 (\omega)
\exp \left \{ \int_{Q_0^2}^{Q^2} \: \frac{dq^2}{q^2} \:
\overline{\gamma} (\overline{\alpha}_S (q^2)/\omega) \right \}
\label{a46}
\end{equation}
\noindent where $\hat{H}^0$ includes the integration constant.
The equality (\ref{a46}) is obtained by integrating the integral
on the left-hand-side by parts.  Thus the saddle-point estimate
of
(\ref{a32}) gives
\begin{equation}
\frac{\partial F_2 (\omega, Q^2)}{\partial \log Q^2} \sim
\overline{\alpha}_S (Q^2) \: \frac{\Phi (\omega,
\overline{\gamma})}{\sqrt{- \overline{\gamma}^2
\tilde{K}^\prime}} \: \exp \left \{ \int_{Q_0^2}^{Q^2}
\frac{dq^2}{q^2} \: \overline{\gamma} (\overline{\alpha}_S
(q^2)/\omega) \right \}
\label{a47}
\end{equation}
\noindent where
$$
\tilde{K}^\prime \equiv \left . d \tilde{K}/d \gamma \right
|_{\gamma = \overline{\gamma}} \quad \hbox{and} \quad
\overline{\gamma} \equiv \overline{\gamma} (\overline{\alpha}_S
(Q^2)/\omega).
$$
This representation is applicable in the region $\omega >
\omega_L \equiv (4 \log 2) \overline{\alpha}_S (Q_0^2)$.  For
smaller values of $\omega$ the saddle-point estimate involves two
(stationary phase) contributions which lead to a different
representation of the integral.  In other words (\ref{a47}) is
not a valid approximation of the integral (\ref{a32}) for $\omega
< \omega_L$.  Unlike the case of fixed $\alpha_S$, the BFKL
solution for running $\alpha_S$ does not contain the branch point
singularity at $\omega = \omega_L$, but rather it has (an
infinite number of) poles in the $\omega$ plane.  The poles are
determined by the starting point condition and hence are
controlled by $\overline{\alpha}_S$ at $Q_0^2$.  It turns out
that the leading pole singularity occurs at $\omega < \omega_L
(\alpha_S (Q_0^2))$ \cite{MKS,LEV2}.

In summary, we have confronted the collinear-factorization
approach for the calculation of observable quantities at small
$x$ with the evaluation based on the $k_T$-factorization formula.
For {\it fixed} $\alpha_S$ both approaches are equivalent at the
leading-twist level.  In fact the insertion of the solution of
the BFKL equation into the $k_T$-factorization formula provides a
recipe for calculating $\overline{\gamma} \equiv \gamma_{gg}$ and
$P_{qg}/\alpha_S$ as power series in
$\overline{\alpha}_S/\omega$, where $\omega$ is the moment index.
The effect of introducing a {\it running} $\alpha_S$ in the
collinear-factorization formalism is summarized by (\ref{a23})
and (\ref{a24}).  We noted that the leading singularity in the
$\omega$ plane is a branch point at $\omega = \omega_L (Q_0^2)$
which is controlled by $\alpha_S (Q_0^2)$ rather than $\alpha_S
(Q^2)$, c.f. (\ref{a27}).  That is the truly perturbative
behaviour is hidden behind a non-perturbative contribution.  In
principle, the latter could be factored off and absorbed into the
starting distribution.  We then examined $k_T$-factorization with
BFKL input with running $\alpha_S$ and compared the predictions
with those obtained from collinear factorization with running
$\alpha_S$.  We found the remarkable result that both the
factorization prescriptions generate a perturbative expansion as
a power series in $\overline{\alpha}_S (Q^2)/\omega$ with exactly
the {\it same} first few non-trivial terms, on top of the same
DLL contribution.  These terms are the most important
perturbative contributions for the onset of the leading $\log
(1/x)$ behaviour in the HERA regime.  In practice, in both the
collinear- and $k_T$-factorization approaches, the leading
singularity in the $\omega$-plane, which controls the small $x$
behaviour, depends on $Q_0^2$.  The location of the singularity
is different, however.  In the first case the power series in
$\overline{\alpha}_S (Q_0^2)/\omega$ builds up a branch point at
$\omega = \omega_L (Q_0^2)$, whereas in the second case we
generate a leading pole at a considerably smaller value of
$\omega$.  We conclude that the truly perturbative contributions
in the two approaches are remarkably similar, but in practice
they are partially hidden by non-perturbative terms.

\bigskip
\noindent {\large \bf Acknowledgements}

We thank J.\ R.\ Forshaw, K.\ Golec-Biernat, R.\ G.\ Roberts, P.\
J.\ Sutton and R.\ Thorne for discussions.  J.K.\ thanks the
Department of Physics and Grey College of the University of
Durham for their warm hospitality.  This work has been supported
in part by the UK Particle Physics and Astronomy Research
Council, Polish KBN grant 2 P302 062 04 and the EU under
contracts no.\ CHRX-CT92-0004/CT93-0357.

\bigskip
\noindent {\large \bf Figure Caption}
\begin{itemize}
\item[Fig.\ 1] Pictorial representation of the
$k_T$-factorization formula of (\ref{a1}).
\end{itemize}

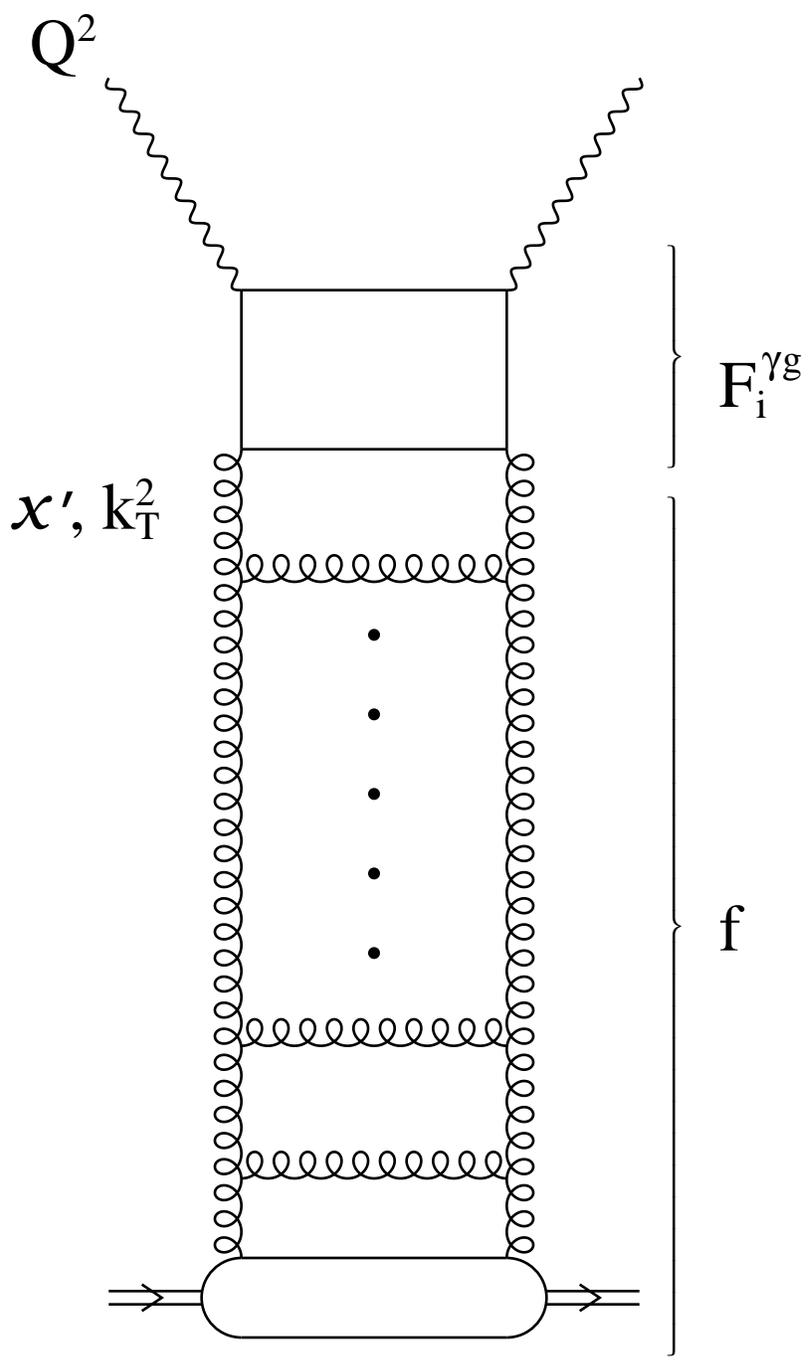
\end{document}